\documentclass{PoS}
\usepackage{booktabs}
\usepackage{multicol}

\title{Jets in QCD matter: Monte Carlo approaches}

\ShortTitle{Jets in QCD matter: Monte Carlo approaches}

\author{\speaker{Liliana Apolinário}\\
        LIP, Avenida Professor Gama Pinto, 2, 1649-003 Lisboa, Portugal\\
        E-mail: \email{liliana@lip.pt}}

\abstract{Monte Carlo approaches are a powerful tool in collider physics as they allow to make theory-data comparison on complex multi-particle observables, otherwise difficult for perturbative calculations. In heavy-ion collisions, there is a multitude of Monte Carlo approaches that try to address jet quenching phenomena, name given to the collection of medium-induced modifications that high momentum particles and jets undergo when traversing the hot and dense medium that is produced in such collisions. These models are being continuous developed alongside the theoretical efforts to understand and accurately describe experimental results provided by both RHIC and the LHC. In this manuscript, it is given a general overview about the fundamental building blocks that these tools have to address to describe jets in heavy-ion collisions. It follows a comparison on the latest results provided by some of the jet quenching Monte Carlo models to jet and intra-jet observables. A final outlook is presented at the end of the manuscript.}

\FullConference{International Conference on Hard and Electromagnetic Probes of High-Energy Nuclear Collisions\\
		30 September - 5 October 2018\\
		Aix-Les-Bains, Savoie, France}

\begin{document}

% ------------------------------------------------------------
\section{Introduction}

\par In heavy-ion collisions the parton shower is a complex multi-scale process. Partons interact with the produced quark-gluon plasma (QGP) and in addition to vacuum-like splittings, soft to semi-hard partons are produced in medium-induced gluon radiation. Moreover, due to the passage of the produced jet, medium constituents become correlated with the jet, a process known as medium response. These correlated soft particles will participate in the jet clustering and are therefore relevant to retrieve information about the medium properties.
\par On top of being a multi-scale object, a parton shower is a space-temporal evolving structure, which adds to the complicated picture of modelling the parton shower evolution. To tackle such a difficult problem there are several approaches that can be grouped into (i) analytical (including perturbative and non-perturbative methods) and (ii) Monte Carlo (MC) event simulations. The former have the advantage of being based on first principle calculations that address elementary processes at parton level. Due to this fact, they have to be carried out observable-by-observable, and the results might be difficult to compare with complex multi-particle observables. Nonetheless, they provide a clearer interpretation of the observations. There have been several advances along this line~\cite{Mehtar-Tani:2014yea,CasalderreySolana:2012ef,Apolinario:2014csa}, in particular on a consistent treatment of vacuum-like and medium-like emissions~\cite{Caucal:2018dla}. However, QGP response is still difficult to be accommodated within this picture. 
\par MC approaches, on the other hand, are based on a factorisation of processes and scales, allowing to build a tool for theory-data comparison on a broader set of observables. While this approach can go beyond analytical calculations, as it can contain the full jet shower evolution on top of an evolving medium, it is still based on theoretical models. As such, MC tools not only inherit the limitations of the theoretical model they are based on, but they also contain further phenomenological assumptions to go beyond the validity region of the analytical approach. For this reason, in the last years, several different jet quenching MC models (see, e.g: \cite{Armesto:2009fj,Zapp:2013vla,KunnawalkamElayavalli:2017hxo,Schenke:2009gb,Park:2018acg,Cao:2016gvr,He:2015pra,Tachibana:2017syd,Casalderrey-Solana:2014bpa,Casalderrey-Solana:2016jvj,Hulcher:2017cpt,Lokhtin:2011qq,Cao:2017qpx,Cao:2017zih}) were put forward to tentatively describe, in a consistent way, jet quenching observables. 
%mong the most recent developments, there is the JETSCAPE collaboration, a modular framework that now includes four previous models (MARTINI, MATTER, LBT and the Hybrid strong/weak coupling approach). 
The reasons for such a multitude of models reside in the several ways one can address: parton shower, medium-induced energy loss, collisional energy loss, medium recoils, medium evolution, hadronization, initialization, ...
%\begin{multicols}{2}
%\begin{itemize}
%	\item parton shower;% as a full in-medium dynamic parton shower or as an afterburner;
%	\item medium-induced energy loss;
%	\item collisional energy loss;
%	\item medium recoils;
%	\item medium evolution; %: as Bjorken expansion, a 3 dimensional hydrodynamic medium or as a strongly coupled fluid;
%	\item hadronization; %: taken as in vacuum or including medium effects;
%	\item initialization. %(spatial distribution of the hard scattering and the moment that the outgoing particles start to interact with the medium)
%\end{itemize}
%\end{multicols}
With all these unconstrained ingredients, the large number of phenomenological tools that are available is no surprise. In this manuscript, instead of detailing each MC model (we refer the reader to the corresponding references instead), we will go through most of the items mentioned above
%\footnote{We will skip medium recoils as they are carefully discussed in \cite{T.Luo}} 
to show, in a more qualitative way, what are the basic ingredients that each MC model tries to implement and the available choices when attempting to build such a tool. This will be the content of section \ref{sec:2}. In section \ref{sec:3}, we will discuss comparisons of MC results with data. Finally, a personal outlook and conclusions are presented in section \ref{sec:4}.

% ------------------------------------------------------------
\section{Jet Quenching Monte Carlo: what do we need?}
\label{sec:2}

\par Most of Jet Quenching MC models are modifications of a preexisting Monte Carlo proton-proton (pp) event generator, usually Pythia 6~\cite{Sjostrand:2006za} or Pythia 8~\cite{Sjostrand:2014zea}. After the generation of the pp collision, the final state partonic shower is modified to accommodate medium-induced modifications. In addition, some of them also add additional soft particles to account for medium-response and jet-induced effects (see \cite{T.Luo}).
\par Energy loss processes based on single gluon emission can be easily implemented within perturbative QCD based MC models. Most of them are based on single-gluon emission calculations that were initially carried out in the eikonal approximation: BDMPS-ASW~\cite{Baier:1996kr,Zakharov:1998sv}, AMY~\cite{Arnold:2001ba}, Higher-Twist~\cite{Guo:2000nz} or GLV~\cite{Gyulassy:2000fs}. While each of these formulations have their own kinematic approximations, they were shown to agree fairly well within their validity region. The extrapolations needed to cover the full kinematic region and full parton shower might provide significantly different results though~\cite{Armesto:2011ht}. 
\par
\begin{figure}[h!]
  \includegraphics[width=\linewidth]{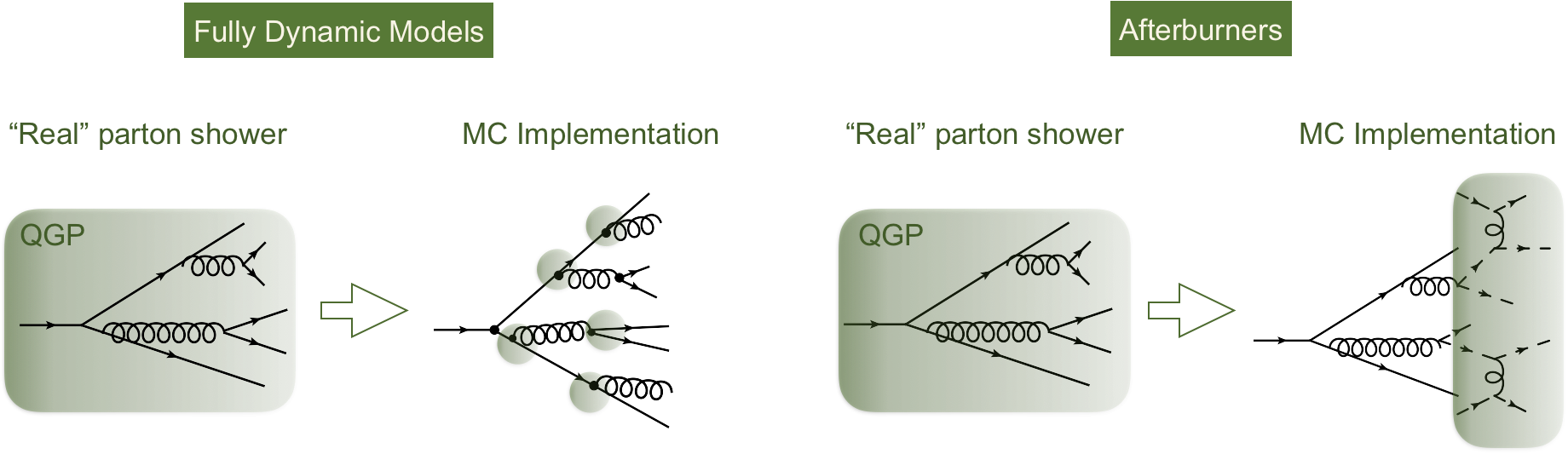}
  \caption{Illustration of the locations where medium-induced effects take place on a vacuum parton shower for fully dynamic models (left) and afterburners (right).  Quarks are represented by black lines, gluons by springs and the QGP by a green background. In the fully dynamic models (left) medium-induced modifications are applied to the the vacuum parton shower during the parton branching. In the afterburners, the particles from a (semi-) developed vacuum parton shower are used as an input to estimate the amount of medium-induced effects that were acted on them during the full shower evolution.}
  \label{fig:splitting}
\end{figure}

 However, not only the energy loss model is different. The way it is implemented within each model, i.e, the procedure to change the vacuum parton shower, will also provide a broad range of possible outcomes. Going from single gluon emission to full jet shower still lacks solid theoretical results despite all the efforts so far. For this reason, current MC models rely on pure phenomenological assumptions to implement these effects. In particular, there are two broad categories: the ones that implement medium-induced effects by changing the parton shower evolution; and the ones that behave as \emph{afterburners}, in which the modifications act on a (semi-)developed partonic shower. In the first category we can find, e.g, Q-PYTHIA~\cite{Armesto:2009fj}, that add to the splitting probability an equivalent in-medium radiation probability from the BDMPS-ASW gluon radiation spectrum. This contribution does not have the same double logarithmic divergences as the vacuum splitting probability ($P_{vac}$), but it is assumed that it can be added as a small correction into the Sudakov form factor. Thus, the total splitting probability of emitting a parton with a fraction of momentum $z$ of the parent one can be written as, 
$P^{tot}(z) = P^{vac}(z) + \Delta P^{med} (z)$.
Pictorially, this corresponds to what is illustrated in figure \ref{fig:splitting} (left), where the green background represents the medium. In this case, medium-induced modifications may act during the whole parton shower evolution as they are implemented directly in the splitting probability. MATTER~\cite{Cao:2017qpx} and JEWEL~\cite{Zapp:2013vla,KunnawalkamElayavalli:2017hxo}, despite the differences, also fall in this category. The \emph{afterburner} type models, like e.g., MARTINI~\cite{Schenke:2009gb,Park:2018acg}, apply AMY rate equations to a vacuum shower that was interrupted at a given momentum scale. Others, like PyQUEN~\cite{Lokhtin:2011qq}, use the results of medium-induced energy loss integrated over the full medium length to rescale the momentum of the partons before hadronization. In this way, they are able to mimic medium-induced modifications that took place during the whole parton shower evolution. Along this line, we can also find Co-LBT~\cite{He:2015pra}, Hybrid strong/weak coupling approach~\cite{Casalderrey-Solana:2014bpa,Casalderrey-Solana:2016jvj,Hulcher:2017cpt}, and others. An illustration for this type of models can be found in figure \ref{fig:splitting} (right).
\par Since the seminal works that first addressed energy loss processes of a parton when going through a colorful medium, there have been several improvements that evolved qualitatively the theoretical picture of a medium-modified jet. In particular, coherence effects~\cite{Mehtar-Tani:2014yea,CasalderreySolana:2012ef}  were shown to be preserved for all the intra-jet structures whose transverse scale was smaller than the transverse resolution of the medium. These structures would then radiate as a single charge in the medium, and their angular pattern would follow an anti-angular ordering, as opposed to vacuum radiation. So far, coherence effects were only implemented in the Hybrid approach (see \cite{Hulcher:2017cpt}), yielding no effect on the resulting jet modifications (see section \ref{sec:3}).

\par In addition to radiative energy loss, partons can also undergo elastic scattering processes. Although not implemented in all MC models, perturbative QCD-based approaches use $2\rightarrow2$ partonic cross-sections ($\sigma$), most at leading order, regularised by a Debye mass $\mu_D$:
%\begin{equation}
$\frac{d\sigma}{d\hat{t}} \simeq \frac{C_R 2 \pi \alpha_s^2}{ (|\hat{t}| + \mu_D^2)^2}$ ,
%\end{equation}
where $\hat{t}$ is momentum transfer, $\alpha_s$ the strong coupling constant and $C_R$ is the Casimir factor in the representation $R$. Only the Hybrid approach uses an AdS/CFT perspective in which the jet loses energy $E$ with distance $x$ as:
\begin{equation}
	\left. \frac{dE}{dx} \right|_{strongly \ coupled} = - \frac{4}{\pi} E_{in} \frac{x^2}{x^2_{stop}} \frac{1}{\sqrt{x_{stop}^2 - x^2}} \, ,
\end{equation}
where $E_{in}$ is the jet energy and $x_{stop}$ the distance over which the jet would lose all of its energy. The later is related to the medium temperature through a fitting parameter, $\kappa_{sc}$. 

\par Another effect that we need to consider in order to have a more realistic estimate of medium-induced effects is the evolution of the medium. 
%The medium is usually modelled by a parameter, $\hat{q}$, that translates the squared average transverse momentum that a particle acquires by mean-free path inside of the medium. It is expected that for an expanding medium this parameter to evolve with time. 
Several options are currently implemented:
\begin{itemize}
	\item PQM model~\cite{Dainese:2004te}, where a 1D nuclei overlap is used to calculate an effective energy loss; in this way the $\hat{q}$ parameter, that translates the squared average transverse momentum that a parton acquires per mean-free path inside of the medium, depends on a single space-time variable; 
	\item Longitudinal Bjorken expansion, that might include also transverse (flow) expansion;
	\item A full hydrodynamic simulation (in most models, the hydrodynamic profiles as a function of time and medium density must be provided by the user).
\end{itemize}
\par Two features, up to now, are common to all MC models: path-length dependence and hadronization. The first is accounted by sampling the production point from the initial nuclei overlap. The second is assumed to be the same as in vacuum (the hadronization model is usually taken from Pythia), but, for the models that allow it, it might include the recoiled particles as well.

%\par A summary of a reduced sample from the available jet quenching MC models can be found in figure \ref{fig:table}.
%\begin{figure}[h!]
%  \includegraphics[width=\linewidth]{table.png}
%  \caption{Partial summary of selected jet quenching Monte Carlo models.}
%  \label{fig:table}
%\end{figure}

% ------------------------------------------------------------
\section{Monte Carlo Results - Data Comparison}
\label{sec:3}

\par
\begin{figure}[h!]
\centering
  \includegraphics[width=\linewidth]{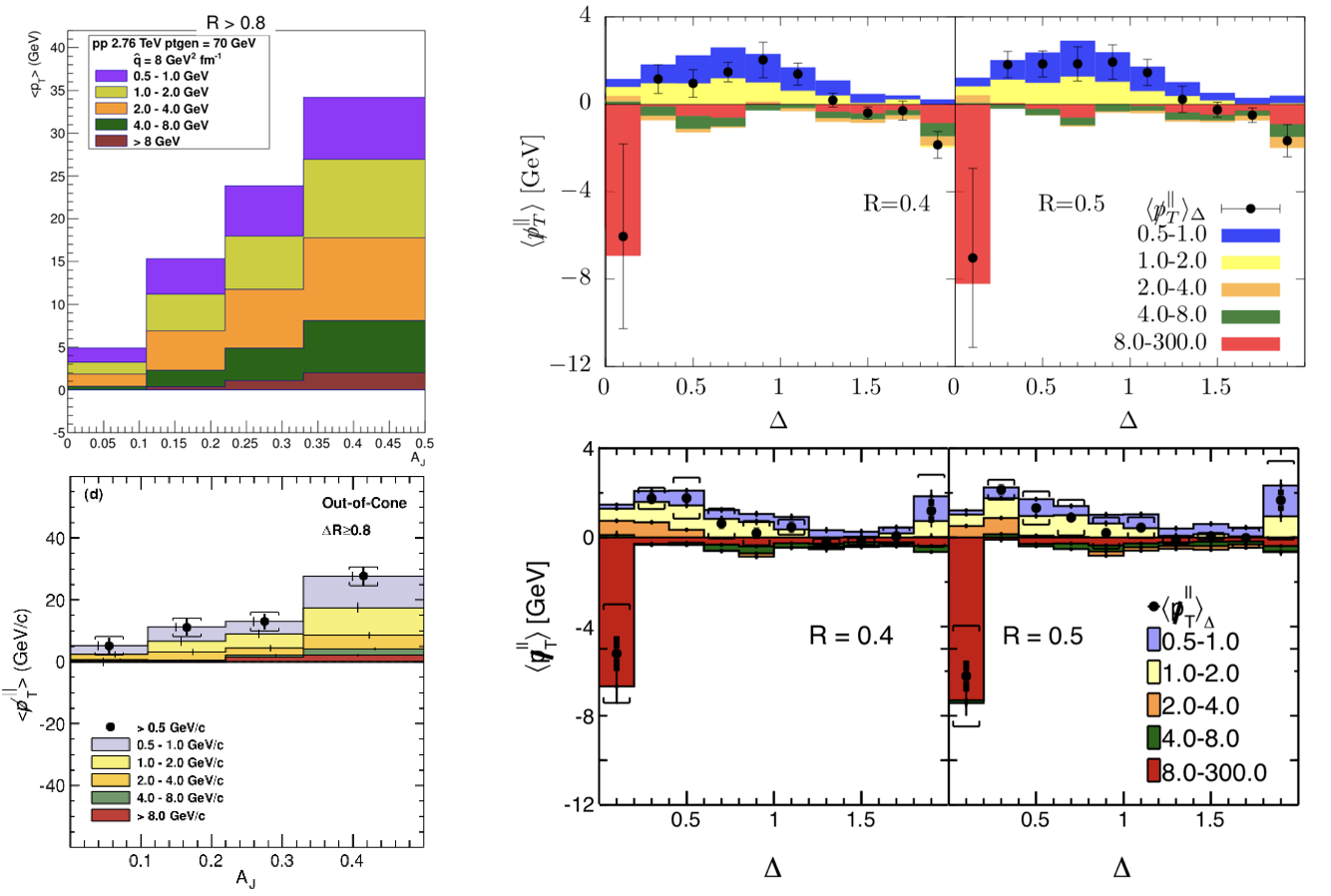}
  \caption{Missing transverse momentum as measured by CMS (lower plots, left~\cite{Chatrchyan:2011sx} and right~\cite{Khachatryan:2015lha}) compared to Q-PYTHIA~\cite{Apolinario:2012cg} (upper left) and Hybrid strong/weak coupling~\cite{Casalderrey-Solana:2016jvj} (upper right)}
  \label{fig:missingpt}
\end{figure}

 When putting forward a more quantitative comparison of jet quenching MC results with experimental data, they seem to perform fairly well within the experimental uncertainties. Nonetheless, a closer look reveals that none of the current models is able to describe, in a consistent way, all jet quenching observations. This naturally leads to the question "\textit{what have we (not) learned so far?}". To answer to this, one can go back to the missing transverse momentum~\cite{Chatrchyan:2011sx,Khachatryan:2015lha} as measured by CMS. The large angle radiation that was observed was soon attributed to medium-induced radiation~\cite{Apolinario:2012cg} (where semi-hard emissions, by subsequent branchings, were observed as soft particles at large angles, see figure \ref{fig:missingpt} (left)). However, in the last couple of years, models that lack medium-induced radiation were able to describe qualitatively the observations through the inclusion of medium response effects. While the agreement is not perfect (see figure \ref{fig:missingpt} right), the resulting transverse momentum spectrum is formed by soft particles mainly located at angular distances larger than the jet cone (R = 0.4 - 0.5 and the $(\eta, \phi)$ distance of a particle $i$ with respect to the jet-axis is defined as $\Delta = \sqrt{ (\eta_{i} - \eta_{jet})^2 + (\phi_{i} - \phi_{jet})^2}$).

\par
\begin{figure}[h!]
\centering
  \includegraphics[width=0.9\linewidth]{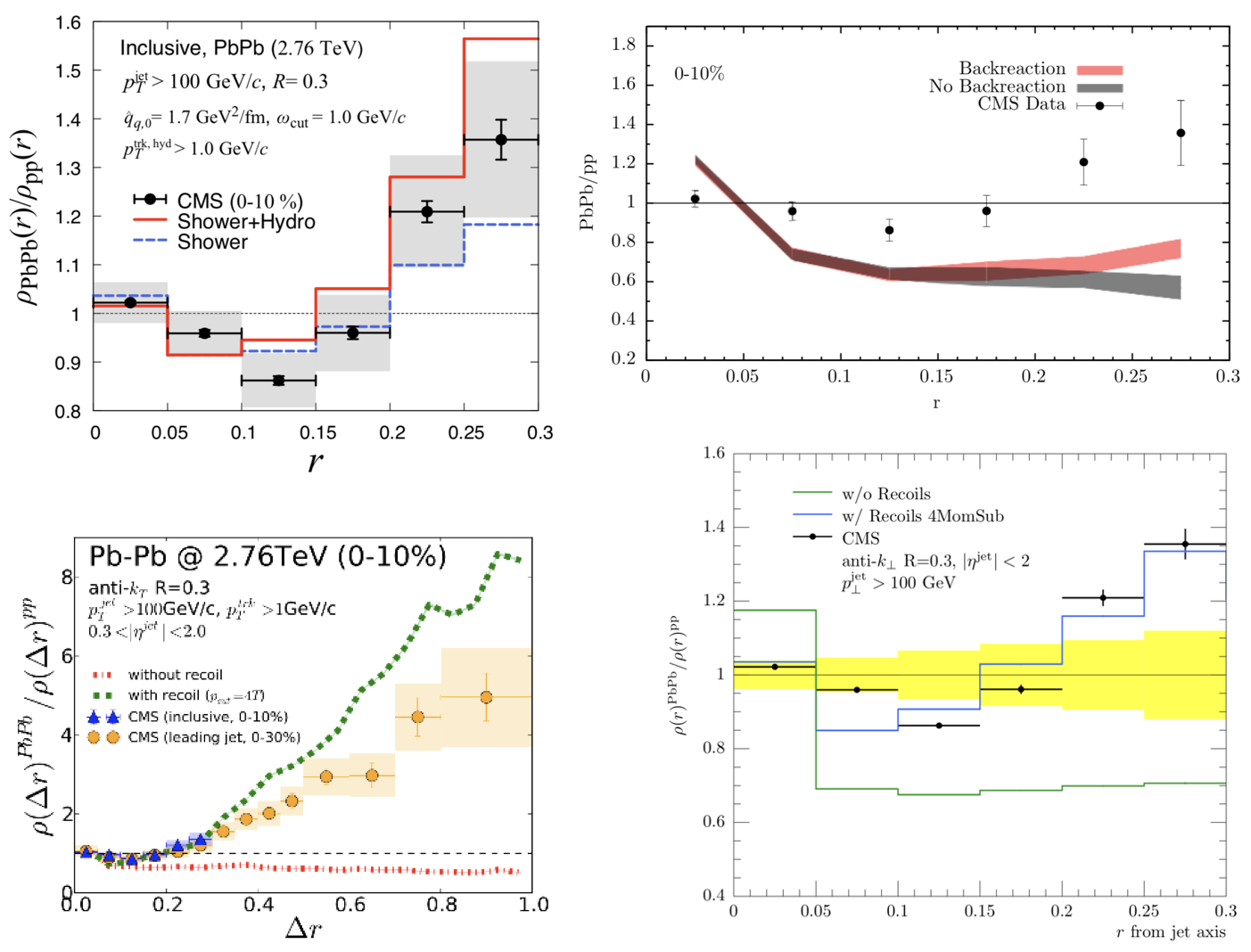}
  \caption{Jet Radial profile as given by Coupled Jet-Fluid~\cite{Tachibana:2017syd} (upper, left), Hybrid~\cite{Hulcher:2017cpt} (upper, right), MARTINI~\cite{Park:2018acg} (lower, left) and JEWEL~\cite{KunnawalkamElayavalli:2017hxo} (lower, right). The CMS data~\cite{Chatrchyan:2013kwa} are also represented as circles in each plot. Both $\Delta r$ and $r$ represent the radial distance to the jet-axis (as the $\Delta$ quantity defined above).}
  \label{fig:jetff}
\end{figure}

 The role of the medium response and its contribution to a particular jet observable is not exclusive to large angles. Measurements of the jet radial profile were one of the smoking guns for the presence of a jet-correlated medium response. It was observed an increase of momentum density at distances within the jet cone, but away from the jet-axis, with respect to a proton-proton baseline. Models with medium-induced radiation are only able to describe this feature if a medium response component is considered. While there are a couple of models that are able to nicely reproduce the experimental data when accounting for medium recoil effects, the magnitude of the recoil component is model dependent (see, e.g. figure \ref{fig:jetff} upper left - Coupled Jet-Fluid analytical approach - vs figure \ref{fig:jetff} lower right - JEWEL MC approach). In addition, it is highly challenging to have a consistent description of the jet radial profile over the whole distance range ($0 < \Delta r < 1.0$, being $\Delta r$ the radial distance with respect to the jet axis), and/or the missing transverse momentum. Models that include a medium response component on top of medium-induced radiation, (see e.g. MARTINI in figure \ref{fig:jetff} lower left) produce an excess of momentum density for $\Delta r > 0.3$ when compared to experimental data. However, the medium response as implemented in the Hybrid model, produces the soft and large angle particles required to describe the missing momentum observable (see figure \ref{fig:missingpt} right), but the magnitude of the effect does not provide a correct description of the jet radial profile within the jet cone (figure \ref{fig:jetff} upper right). This simultaneous description is even more challenging if the jet mass is also considered.

\par Despite the differences, MC approaches with medium-induced radiation seem to predict the same qualitative features as the analytical results. However, the phenomena of jet coherence, as implemented in the Hybrid model, does not seem to correspond to theoretical expectations. It was shown that jet coherence has a significant impact on the jet radial profile~\cite{Mehtar-Tani:2014yea}, by enhancing the momentum away of the jet axis. Furthermore, it was expected that the resulting nuclear modification factor ($R_{AA}$) was also enhanced~\cite{Mehtar-Tani:2017web}. However, results by Hybrid show a very mild effect on the jet radial profile~\cite{Hulcher:2017cpt}, and an almost unchanged jet $R_{AA}$~\cite{Knichel}. The difference between analytical and MC results on $R_{AA}$ might be attributed to the necessary re-fitting to experimental data when introducing coherence effects. This also contributes to the small effect seen on the jet radial profile, but further studies are required to better understand the effect of jet coherence on the in-medium jet development.

% ------------------------------------------------------------
\section{Outlook \& Summary}
\label{sec:4}

\par A systematic comparison of models to the same set of observables is urgently needed. In particular, to understand the role of the medium response in jet observables, one could use the jet shapes with the transverse momentum information as provided by the CMS collaboration~\cite{Sirunyan:2018jqr}. While the magnitude of the jet shape shows already some discrimination power between models, it is necessary to understand the transverse momentum spectrum of the recoil particles, both at small and large distances away from the jet axis. Since these particles are allowed to participate in the hadronization mechanism, they will result in a different final hadro-chemistry that might be further investigated. Measurements of the jet fragmentation and jet transverse profile in Z/$\gamma$-jet events can provide further understanding on the medium response mechanism. This is experimentally challenging due to the large statistics that is required. But if recoils dominate low-energy/wide-angle particle distribution, the result of these jet shapes should change as a function of the boson transverse momentum. 

\par A complementary measurement is to use observables that are insensitive to a particular in-medium jet modifications: e.g, the $\Delta S_{12}$ proposed in \cite{Apolinario:2017qay} and recently measured by STAR~\cite{Raghav}, is less sensitive to medium recoil effects allowing a precise determination of medium-induced radiation; the use of Lund diagrams in heavy-ions~\cite{Andrews:2018jcm}, also measured by ALICE~\cite{Andrews:2018wgw}, allow to isolate phase space regions that might be dominated by medium recoiled particles using Soft-Drop techniques. 

\par So far, all proposed observables assess the final result of medium-induced effects that took place during the full medium (and parton shower) evolution. A promising way of constraining jet-medium interactions and, ultimately, the QGP properties, is to devise a strategy to measure jet quenching effects at different timescales. For instance, using time-delayed probes, as boosted tops~\cite{Apolinario:2017sob}, it is possible to have sensitivity to quenching effects that took place during later times of the QGP lifetime.

\par The variety of jet quenching models, which might seem surprisingly large, is a direct consequence of our theoretical uncertainty. While there is not yet a consistent picture, efforts are being made to further test and constrain MC models; (i) devising new kind of probes, (ii) continuous development of MC modelling and (iii) creation of a useful testing environment that accommodate several different models, like the recent JETSCAPE Collaboration~\cite{Cao:2017zih}. While they cannot be used as simple black-boxes, jet quenching MC models based on theoretical results are a powerful tool to take full advantage of current (RHIC and LHC) and future accelerator data.

% ------------------------------------------------------------
\vspace{3mm}
\textbf{Acknowledgements}: The author would like to thank N. Armesto, R. Concei\c{c}\~{a}o, L. Cunqueiro, G. Milhano, C. Salgado and M. Verweij for the useful comments provided during the elaboration of this talk and to the referee for improving the readability of this manuscript. The author acknowledge the financial support by Funda\c{c}\~{a}o para a Ci\^{e}ncia e Tecnologia (FCT) under CERN/FIS-PAR/0022/2017 and OE - Portugal, FCT, I.P., under DL/57/2016/CP1345/CT0004.

% ------------------------------------------------------------
\bibliographystyle{JHEP}
\bibliography{bibliography}
%\begin{thebibliography}{99}
%\bibitem{...}
%....
%
%\end{thebibliography}

\end{document}